\documentclass[reprint,
superscriptaddress,
showpacs, 
nofootinbib
amsmath,amssymb,
aps,
pra,
]{revtex4-2}
\usepackage{graphicx}
\usepackage{dcolumn}
\usepackage{bm}
\usepackage{dsfont}
\usepackage{physics}
\usepackage{bbold}
\usepackage{color}
\usepackage{soul}
\setstcolor{red}
\usepackage{ifthen}
\usepackage{url}
\usepackage[hidelinks]{hyperref}
\hypersetup{
    colorlinks,
    linkcolor={blue},
    citecolor={blue},
    urlcolor={blue}
}

\usepackage{ifthen}
\newboolean{showannotations}   
\setboolean{showannotations}{true} 
\newcommand*{\added}[1]{%
  \ifthenelse{\boolean{showannotations}}{{\color{black}#1}}{#1}%
}

 \usepackage[thinc]{esdiff}

\newcommand*{\removed}[1]{%
  \ifthenelse{\boolean{showannotations}}{\st{#1}}{}%
}

\newcommand*{\change}[2]{%
  \ifthenelse{\boolean{showannotations}}{{\color{red}\st{#1}}{\color{blue}#2}}{#2}%
}

\begin{document} 

\title{  Exciton-Polariton Dynamics in Multilayered Materials}
\author{Saeed Rahmanian Koshkaki}%
\email{rahmanian@tamu.edu}
\affiliation{Department of Chemistry, Texas A\&M University, College Station, Texas 77843, USA}
\author{Arshath Manjalingal}%
\affiliation{Department of Chemistry, Texas A\&M University, College Station, Texas 77843, USA}
\author{Logan Blackham}
\affiliation{Department of Chemistry, Texas A\&M University, College Station, Texas 77843, USA}
\author{Arkajit Mandal}%
\email{mandal@tamu.edu}
\affiliation{Department of Chemistry, Texas A\&M University, College Station, Texas 77843, USA}

\begin{abstract}
\footnotesize
Coupling excitons with quantized radiation has been shown to enable coherent ballistic transport at room temperature inside optical cavities. Previous theoretical works employ a simple description of the material, depicting it as a \added{one-dimensional} single layer placed in the middle of an optical cavity, thereby ignoring the spatial variation of the radiation \added{field}. In contrast, in most experiments, the optical cavity is filled with organic molecules or \added{multiple layers of two-dimensional} materials. Here, we develop an efficient mixed-quantum-classical approach, introducing a {\it bright layer} description, to simulate the exciton-polariton quantum dynamics \added{in three dimensions}. Our simulations reveal that, for the same Rabi splitting, a multilayered material extends the quantum coherence lifetime and enhances transport compared to a single-layer material. We find that this enhanced coherence can be traced to a synchronization of phonon fluctuations over multiple layers, wherein the collective light-matter coupling in a multilayered material effectively suppresses the phonon-induced dynamical disorder.
\end{abstract}

\maketitle
{\footnotesize

{\bf Introduction.}  Quantum coherence in the condensed phase typically lasts tens of femtoseconds at room temperature, due to phonon-induced decoherence intrinsic to the material itself~\cite{Nitzan2006}. Recent experiments have demonstrated that light-matter interaction inside optical cavities enables coherent quantum propagation of exciton-polaritons (EP) by effectively shielding polaritons from phonon fluctuations~\cite{XuNC2023, BalasubrahmaniyamNM2023, KeelingNM2023, SandikNM2024, PandyaAS2022, MandalCR2023, blackham2025}. Specifically, prior theoretical works~\cite{XuNC2023, TichauerAS2023, SokolovskiiNC2023,KruppArxiv2024, YingArxiv2024,KruppArxiv2024}, show that polaritons, which are part light and part matter, couple to phonons more weakly compared to bare excitons because only their matter component couples to the phonons, which is partly responsible for the sustained coherence lifetime of exciton-polaritons. In addition to this, our recent theoretical work suggests that the light-matter interactions approximately restore the phonon-induced translational symmetry breaking in materials~\cite{blackham2025}. Overall, these works reveal that coupling cavity radiation to materials provides tuning knobs for controlling phonon-induced decoherence beyond the traditional paradigms of material synthesis, paving the way for developing next-generation quantum devices~\cite{LiewOME2023, XiangCR2024}. 

However, these theoretical works adopt a simplified \added{light-matter system, where} a single layer, \added{modeled as an one-dimensional chain}, placed at the center of \added{a two-dimensional} optical cavity. This is in contrast to the experiments~\cite{XuNC2023,BalasubrahmaniyamNM2023, PandyaAS2022, SandikNM2024, MandalNL2023, BalasubrahmaniyamPRB2021} where the space between the reflective mirrors is filled with molecules or multiple layers of \added{two-dimensional} materials, as schematically illustrated in Fig.~\ref{Fig1_schemband}a-b. Consequently, the density of states in these two setups (i.e., multilayered vs. single-layered) is drastically different (see insets in Fig.~\ref{Fig1_schemband}c-d), \added{with a} multilayered material featuring an ensemble of optically dark bands that are absent in a single-layered material when coupled to a cavity. Therefore, it is anticipated that the polariton quantum dynamics in these two setups will differ. However, the role of a multilayer configuration in polaritonic quantum dynamics has remained unexplored, which is the focus of this present work.


\begin{figure} 
\centering
\includegraphics[width=1.0\linewidth]{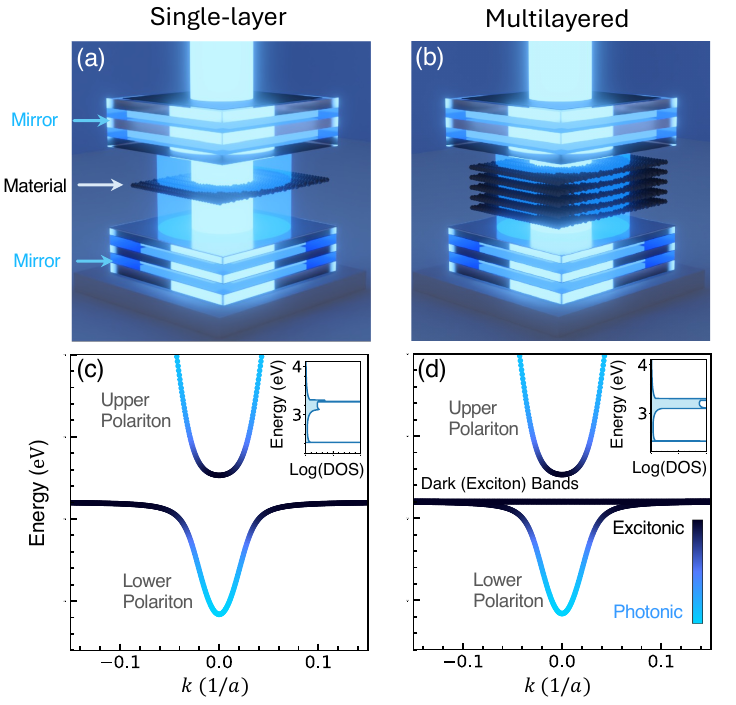}
\caption{\footnotesize \textbf{ Schematics and band structures of single-layer vs. multilayer materials.}  Illustration of (a) a single-layer material and (b) a multilayered material in \protect\added{ a three-dimensional} optical cavity, respectively.  The exciton-polariton band structures of (c) single-layer and (d) multilayered materials in the absence of phonons ($\gamma=0$) \protect\added{in a two-dimensional optical cavity}. In a multilayer setup, in addition to the polariton bands, an ensemble of dark (material) bands exists. These dark bands increase the density of states (DOS) near the onsite energy value, which are plotted as insets within (c)-(d). }
\label{Fig1_schemband}
\end{figure}

Simulating exciton-polariton dynamics in multilayered material is an extremely challenging task, as the number of required unit cells in a single layer exceeds $\approx 10^{4} - 10^8$, thus requiring us to perform a quantum dynamics simulation involving $\approx 10^{6} - 10^{10}$ states for $\approx 10^2$ layers, typical in these systems. To resolve this fundamental challenge, here we introduce an efficient mixed quantum-classical approach that uses a \textit{bright-layer unitary transformation} of the light-matter interaction and propagates the dynamics utilizing a split-operator approach. \added{We emphasize, that this approach do not introduce any additional approximation.} Using this convenient theoretical tool, we investigate the polaritonic quantum dynamics in multilayered materials. We discover a {\it phonon fluctuation synchronization effect} where the exciton-polaritons experience much lower phonon fluctuations and remain coherent for a longer time. We show that this effect originates from the delocalized nature of light-matter interactions, leading to an averaging of phonon fluctuations and suppression of phonon-induced dynamical disorder. Here, we show that multilayered materials can extend the exciton-polariton coherence lifetime, up to an order of magnitude, and enhance exciton-polariton transport, which is relevant for developing polariton-based quantum devices~\cite{SandikNM2024, XiangCR2024, SanvittoNM2016}. 


{\bf Theory.} To explore the dynamics of exciton-polaritons formed in multilayered materials, we consider a light-matter Hamiltonian beyond the long-wavelength approximation~\cite{MandalNL2023,MandalCR2023, TaylorCPR2025,KeelingARPC2020, tichauerJCP2021}. \added{This Hamiltonian} can be derived from the non-relativistic quantum electrodynamics Hamiltonian in the Coulomb gauge~\cite{MandalNL2023,JiajunPRB2020}, written as (in atomic units)
\begin{align}\label{RealSPaceHamiltonian}
\hat{H}_\mathrm{LM} = \hat H_\text{ex} + \hat H_\text{phn}  + \hat H_\text{cav} + \hat H_\mathrm{ex-phn} +  \hat H_\text{ex-cav} + \hat H_\text{loss}\, ,
\end{align}
where $\hat H_\text{ex}$, $\hat H_\text{phn}$ and $\hat H_\text{cav}$
are the bare excitonic, phonon, and cavity Hamiltonians, with $\hat H_\mathrm{ex-phn}$ and $\hat H_\text{ex-cav}$ describing the exciton-phonon and exciton-cavity interactions. \added{Note that, for the sake of simplicity, we first restrict our discussion to a simple stacked 1D layered system (extending in $x$ and $z$ directions), as we arrive at the same conclusions when considering the full three dimensional problem as shown in Fig.~\ref{Fig5_3Dcavity}. The full three dimensional Hamiltonian is provided in SI. Also note that, as schematically illustrated in Fig.~\ref{Fig1_schemband}, we consider the cavity mirrors to be distributed Bragg reflectors (DBR), so we can ignore the formation of surface plasmon polaritons, which exist at metal-dielectric interfaces~\cite{maier2007plasmonics}.} Here, the last term $\hat H_\text{loss}$ describes cavity photon loss. The bare excitonic Hamiltonian is \added{a one-dimensional tight-binding model with one exciton per unit cell, which is given by}
\begin{align}\label{exciton}
\hat H_\text{ex} = \sum_{n}^{N} \sum_{m}^{M} \left[\epsilon_0 \hat  X^\dagger_{n, m}\hat X_{n,m} - \tau     \left(\hat X^\dagger_{n+1, m}\hat X_{n, m} + h.c.\right)  \right]\, , 
\end{align}
where $\hat{X}^\dagger_{n, m}$ creates an exciton at the site $n$ in the $m$th layer, $ \epsilon_0$ is the onsite energy and $\tau$ is the nearest-neighbor hopping parameter. These layers are stacked parallel to each other with an interlayer spacing of $a_z = 4$ nm, with each site separated laterally along the $x$ direction by a distance of $a = 1.2$ nm, which corresponds to typical perovskite materials~\cite{Svenja2020FrenkelHolsteinperovskites}. \added{Therefore, the spatial location of the exciton $\hat{X}^\dagger_{n, m}$ is $R_{n,m} = x_n \vec{e}_x + z_m \vec{e}_z = na \vec{e}_x + ma_z \vec{e}_z$.} In this work, we also consider one phonon degree of freedom per site, with the bare phonon Hamiltonian given by 
\begin{align}\label{phonon}
\hat{H}_\text{phn} = \sum_{n,m}\frac{1}{2}\left({\hat{p}_{n,m}^2}+\omega^2 \hat{q}_{n,m}^2  \right),
\end{align}
where $\hat{p}_{n,m}$ and $\hat{q}_{n,m}$ are the momentum and position of the phonons with frequency $\omega=1440~\text{cm}^{-1}$~\cite{Svenja2020FrenkelHolsteinperovskites}. We consider a typical form of the exciton-phonon coupling~\cite{XuNC2023,troisi2006charge,ArnardottirPRL2020,terry2023theory} described by 
\begin{align}\label{exc-phonon}
\hat H_\mathrm{ex-phn} = \gamma \sum_{n, m}  \hat{q}_{n,m} \hat X^\dagger_{n, m}\hat X_{n,m},
\end{align}
where $\gamma$ is the exciton-phonon coupling strength.  The bare cavity Hamiltonian $\hat H_\text{cav}$ describes a set of confined radiation modes, \added{considering only transverse magnetic (TM) polarized mode (we consider both polarizations in the full three dimensional scenario in Fig.~\ref{Fig5_3Dcavity}),} in a Fabry-P\'{e}rot optical cavity~\cite{LitinskayaPSS2004, MandalCR2023, KockumNRP2019, WeiPR2024} such that 
\begin{align}\label{photon}
\hat H_\text{cav} = \sum_{\boldsymbol{k} }\omega_{\boldsymbol{k}} \hat a^\dagger_{\boldsymbol{k}} \hat a_{\boldsymbol{k}}\, ,
\end{align}
where $\hat a^\dagger_{\boldsymbol{k}}$ creates a photon of wavevector ${\boldsymbol{k}}$ with a frequency $\omega_{\boldsymbol{k}} = \frac{c}{\eta} |{\boldsymbol{k}}|$ where $c$ and $\eta  = 2.4$ are the speed of light and the refractive index, respectively. \added{As noted earlier, in the simplified scenario,} we consider two directions $x$ \added{(longitudinal)} and $z$ \added{(transverse)}, such that ${\boldsymbol{k}} = k_x \vec{e}_x + k_z \vec{e}_z$ with $z$ as the cavity quantization direction. Similar to recent work~\cite{tichauerJCP2021, MandalNL2023, ArnardottirPRL2020, ChngNL2025, XuNC2023, TichauerAS2023, JasrasariaJCP2025}  we impose a periodic boundary condition in the $x$ direction, effectively quantizing the $k_x = \frac{2\pi n_x}{N a}$ where $n_x = 0, \pm 1, \pm 2, ...$. \added{Here,} we consider first cavity \added{quantization} mode along the $z$ direction such that $k_z = \frac{\pi}{L}$ with $L = 100$ nm as the distance between the two reflective mirrors of the optical cavity. To simplify our notation we denote $k = k_x$ and label all photonic operators and related parameters with $k$, as $k_z$ is fixed, such that $\hat H_\text{cav} = \sum_{\boldsymbol{k} }\omega_{\boldsymbol{k}} \hat a^\dagger_{\boldsymbol{k}} \hat a_{\boldsymbol{k}} \rightarrow \sum_{{k} }\omega_{{k}} \hat a^\dagger_{{k}} \hat a_{{k}}$ and $\omega_k = \frac{c}{\eta} \sqrt{k_z^2 + k^2}$. 

The light-matter interactions beyond the long-wavelength approximation \added{in the simplified two-dimensional scenario}~\cite{XuNC2023, JasrasariaJCP2025, keeling2020bose, ArnardottirPRL2020} is described by
\begin{align}\label{exciton-cav}
\hat H_\text{ex-cav} = \sum_{n,m,{k}} \frac{\Omega_k}{\sqrt{N}}  \left( \hat X_{n, m}^\dagger\hat a_{k}e^{i{k} x_{n}} + h.c.  \right)\sin(k_z  z_{m})\, ,
\end{align}
where $\Omega_k =  \sqrt{\frac{\omega_0}{\omega_{k}}}  \Omega_0$ is the light-matter coupling strength. \added{Throughout the manuscript we use $\Omega_0 = 330$ meV, therefore the light-matter coupling remains within the strong coupling regime and away from the ultra-strong coupling regime. Other data for larger and smaller $\Omega_0$ can be found in the Supplementary Information (SI)}. Finally,  we model the cavity photon loss using a non-Hermitian Hamiltonian~\cite{tichauerJCP2021,Antoniou2020JPCL, Qiu2021JPCL, KoesslerJCP2022} given by
\begin{align}\label{loss}
\hat H_\text{loss}  = -\frac{i}{2t_c} \sum_k\hat a^\dagger_{{k}} \hat a_{{k}}\, ,
\end{align}
where $t_c$ is the cavity photon lifetime.

\begin{figure*}
\centering
\includegraphics[width=1.0\linewidth]{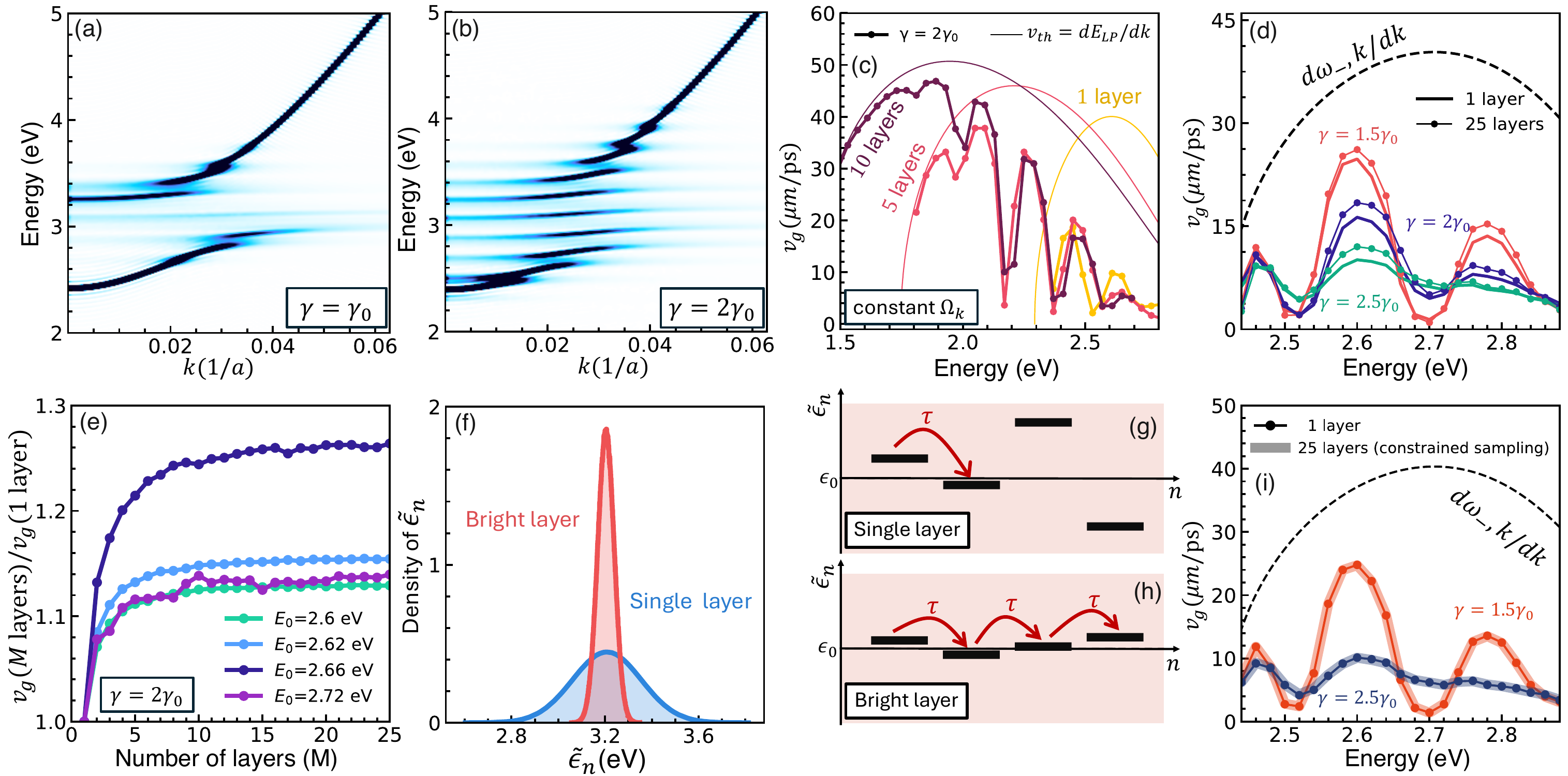}
\caption{\footnotesize \textbf{Exciton-polariton dynamics and spectra in multilayered material ($M=25$) and the phonon fluctuation synchronization effect \protect \added{in a two-dimensional cavity}.} The exciton-polariton angle-resolved spectra at (a) $\gamma=\gamma_0$ and (b) $\gamma=2\gamma_0$. Exciton-polariton group velocity computed from quantum dynamical simulations vs. theoretical curve with (c) light-matter coupling per layer $\Omega_k$ a constant and (d) the total Rabi-splitting $\sqrt{\mathcal{S}}\Omega_k$ a constant. Here, the phonon coupling $\gamma$ is written in terms of $\gamma_0$, a constant. (e) The ratio of group velocities in a multilayered material to a single-layer material. (f)  Distribution of on-site excitonic energy.  Illustration of hopping dynamics in (g) a single-layer material with higher disorder and (h) in a multilayered material with effectively lower disorder.  (i) Group velocity in single-layered and multilayered material with constrained sampling of ${q}_{n,m}(t = 0) = {q}_{n,1}(t = 0)$. ($\gamma_0\approx 5.8\times 10^{-4}$ a.u., \protect \added{$\Omega_0 = 330$ meV, and 60000 sites.})}
\label{fig2_vgs}
\end{figure*}

\begin{figure*}[ht]
    \centering
    \includegraphics[width=0.8\linewidth]{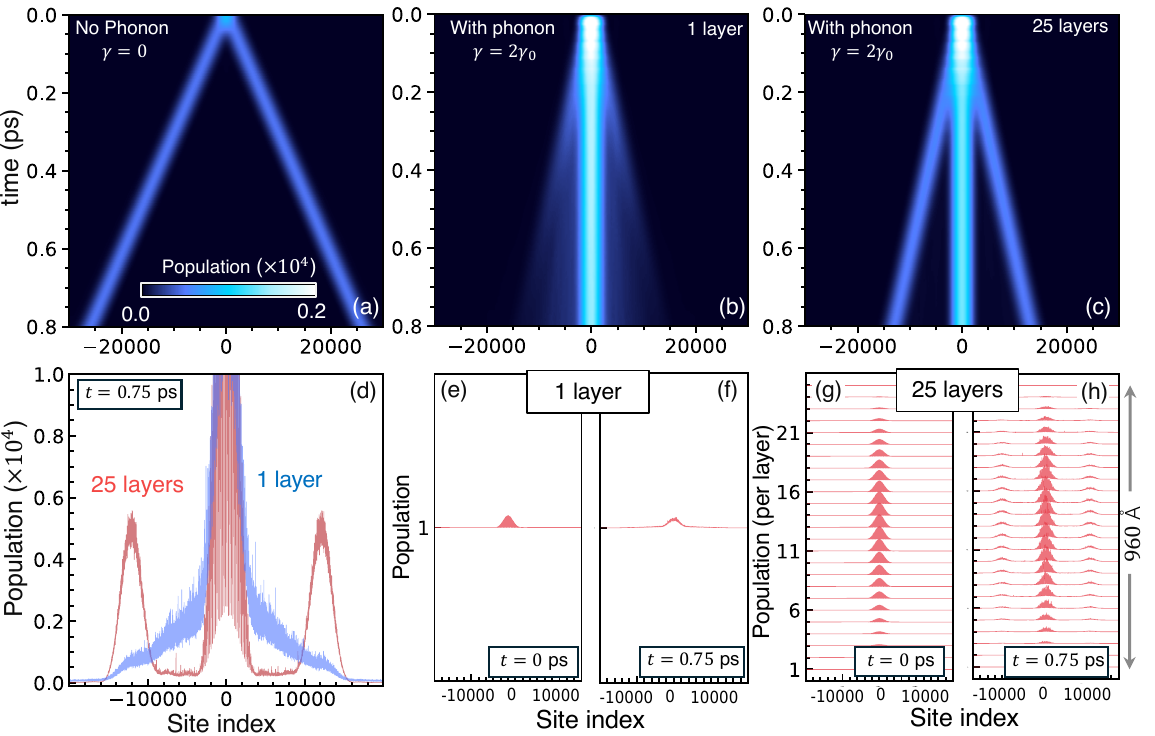}
    \caption{\footnotesize \textbf{Comparing exciton-polariton dynamics in single-layer vs. multilayered materials ($M=25$) \protect \added{in a two-dimensional cavity}  with the initial excitation at $E_0 = 2.62$ eV.}  (a) Time-dependent exciton-polariton population for \protect\added{both single-layer and multilayered material} with no exciton-phonon couplings ($\gamma=0$). (b)-(c) Same as (a) but in the presence of phonon couplings  $\gamma=2\gamma_0$ in (b) a single and (c) multilayered material. (d) Site-resolved exciton density at $t=0.75$ ps and $\gamma=2\gamma_0$. To determine the multilayer exciton populations in (d), we summed the density contributions from all layers at each site. (e)-(h) Site and layer-resolved exciton density for (e)-(f) a single layer and (g)-(h) multilayered material at $t=0$  and $t=0.75$ ps with the same parameters as in (d). ($\gamma_0\approx 5.8\times 10^{-4}$ a.u., \protect \added{$\Omega_0 = 330$ meV, and 60001 sites.}) } 
    \label{Fig3_spread}
\end{figure*}

{\bf Quantum Dynamical Approach.} In this work, we employ a custom mixed-quantum-classical method based on the mean-field Ehrenfest approach to simulate the dynamics of the light-matter hybrid system. Within the standard Ehrenfest approach, that has been extensively used to simulate non-adiabatic dynamics of polaritons,~\cite{Hoffmann2020JCP, Mandal2019JPCL, MandalCR2023, xu2024JCP, LiPRA2018, TichauerAS2023, tichauerJCP2021, SokolovskiiNC2023, akimov2014coherence, alonso2012ehrenfest} the \added{phonon} (or slow) degree of freedom is evolved classically, with $\{\hat{q}_{n,m}, \hat{p}_{n,m}\} \rightarrow \{{q}_{n,m}, {p}_{n,m}\}$, following the Hamiltonian's equation of motion,
\begin{align}
    \dot{p}_{n,m}(t) =  - \Big\langle {\Psi} (t)\Big| \diff{\hat{H}_{\mathrm{LM}}}{q_{n,m}} \Big |{\Psi}(t)\Big\rangle, ~~~\dot{q}_{n,m}(t) = {p}_{n,m}(t),
\end{align}
where $|{\Psi}(t)\rangle$ is the excitonic-photonic wavefunction at time $t$. The excitonic-photonic wavefunction is evolved using the time-dependent Schr\"{o}dinger equation written as 
\begin{align}\label{tdse}
i|\dot{\Psi} (t)\rangle &= \Bigg[\hat{H}_{\mathrm{LM}} - \sum_{n,m}\frac{p_{n,m}^2}{2} \Bigg] |{\Psi (t)}\rangle.
 \end{align}
In this work, we confine the exciton-polariton dynamics to the single excited subspace. \added{Therefore, the exciton-photon wavefunction is given } 
\begin{align}\label{wfn}
|{\Psi} (t)\rangle &= \Big( \sum_{k} c_{k}(t) \hat{a}^{\dagger}_{k} + \sum_{n,m} {b}_{n,m}(t) \hat{X}^{\dagger}_{n,m} \Big) |\bar{0}\rangle \\
&= \sum_{k} c_{k}(t)|1_k\rangle  + \sum_{n,m} {b}_{n,m}(t) |n,m\rangle  
\end{align}
where $c_{k}(t)$ and ${b}_{n,m}(t)$ are time-dependent coefficients. In the second line, we have introduced the compact representation $|1_k\rangle \equiv  \hat{a}^{\dagger}_{k} |\bar{0}\rangle$ and $|n,m\rangle  \equiv \hat{X}^{\dagger}_{n,m}|\bar{0}\rangle $, with $|\bar{0}\rangle$ as the ground (or vacuum) state of the system, for simplicity. 
Despite the mixed quantum-classical treatment of the full light-matter Hamiltonian, numerically solving Eq.~\ref{tdse} is extremely challenging as the present work requires a basis of size $\approx10^{6}$ \added{for the simplified two-dimensional scenario ($\approx10^{8}$ presented in Fig.~\ref{Fig5_3Dcavity})}. To resolve this issue, we develop a {\it split-operator} approach where a short-time (a single time-step) propagation of $|{\Psi} (t)\rangle$ is obtained as
\begin{align}\label{propagate}
|{\Psi} (t &+ \delta t)\rangle =  e^{-i\hat{H}_\mathrm{LM}  \delta t}|{\Psi} (t)\rangle   \\
&\approx  \hat{U}_\mathrm{ft} \hat{U}_\mathrm{B} \Big(\hat{U}_\mathrm{B}^{\dagger} \hat{U}_\mathrm{ft}^{\dagger}  e^{-i\hat{H}_{\mathrm{EP}}\delta t} \hat{U}_\mathrm{ft} \hat{U}_\mathrm{B}\Big) \hat{U}_\mathrm{B}^{\dagger} \hat{U}_\mathrm{ft}^{\dagger}  e^{-i\hat{H}_{\mathrm{env}}\delta t}  |{\Psi} (t)\rangle \nonumber\\
&=  \hat{U}_\mathrm{ft}   \hat{U}_\mathrm{B} e^{-i( \hat{U}_\mathrm{B}^{\dagger}\hat{U}_\mathrm{ft}^{\dagger} \hat{H}_{\mathrm{EP}}\hat{U}_\mathrm{ft} \hat{U}_\mathrm{B})\delta t}   \hat{U}_\mathrm{B}^{\dagger}\hat{U}_\mathrm{ft}^{\dagger}  e^{-i\hat{H}_{\mathrm{env}}\delta t}  |{\Psi} (t)\rangle, \nonumber
\end{align}
where $\hat{H}_\mathrm{LM} = \hat{H}_{\mathrm{EP}} + \hat{H}_{\mathrm{env}}$ with $\hat{H}_{\mathrm{env}} = \hat H_\text{phn} + \hat H_\mathrm{ex-phn} + \hat{H}_\text{loss}$ and $\hat{H}_{\mathrm{EP}} = \hat H_\text{ex} + \hat H_\text{cav} + \hat H_\text{ex-cav}$. Here $\hat{H}_{\mathrm{env}}$ is the diagonal in the exciton-photon basis $\{|1_k\rangle, |n,m\rangle  \}$ chosen here. As a result, the action of the matrix  $e^{-i\hat{H}_{\mathrm{env}}\delta t}$ on the vector $|{\Psi} (t)\rangle$ 
reduces to a simple Hadamard product between a vector containing the diagonal elements of $e^{-i\hat{H}_{\mathrm{env}}\delta t}$ and $|{\Psi} (t)\rangle$.  Meanwhile, $\hat{U}_\mathrm{ft}^\dagger$ is a unitary operator that Fourier transforms the excitonic subspace within each layer, such that $\hat{U}^\dagger_\mathrm{ft}|{\Psi} (t)\rangle = \sum_{k} c_{k}(t)|1_k\rangle  + \hat{U}^\dagger_\mathrm{ft}\sum_{n,m} {b}_{n,m}(t) |n,m\rangle = \sum_{k} c_{k}(t)|1_k\rangle  +  \sum_{k,m} {b}_{k,m}(t) |k,m\rangle$. We perform this Fourier transformation using the Fast Fourier Transformation (FFT) algorithm, which scales as $N\log N$, resulting in a significant reduction of the computation cost~\cite{tannor2007introduction, nielsen2001quantum, ChngNL2025}. Importantly, here we introduce the unitary operator $\hat{U}_\mathrm{B} = \hat{U}_\mathrm{db} \hat{U}_\mathrm{D}$, where $\hat{U}_\mathrm{db}$  transforms the excitonic subspace into a {\it dark-bright} layers subspace~\cite{MandalNL2023} and $\hat{U}_\mathrm{D}$  diagonalizes this transformed exciton-polariton Hamiltonian. \added{This unitary transformation is given by (a much more involved unitary transformation in case of the full 3D scenario is provided in the SI)}

\begin{align}\label{BFT}
\hat{U}_\mathrm{B}^{\dagger}\hat{U}_\mathrm{ft}^{\dagger} \hat{H}_{\mathrm{EP}}\hat{U}_\mathrm{ft} &\hat{U}_\mathrm{B} = \hat{U}_\mathrm{B}^{\dagger}\big[\hat{U}_\mathrm{ft}^{\dagger} (\hat H_\text{ex} + \hat H_\text{ex-cav})\hat{U}_\mathrm{ft} + \hat H_\text{cav}\big]\hat{U}_\mathrm{B}\nonumber\\
&=\hat{U}_\mathrm{D}^{\dagger} \sum_{k} \Big[ \epsilon_{k}\hat X^\dagger_{k,b}\hat X_{k,b}  + \sqrt{\mathcal{S}} \Omega_k \left( \hat X^\dagger_{k,b} \hat a_k + h.c. \right) \nonumber  \\
&~~~~~~~~~~~+ \omega_k\hat a^\dagger_k \hat a_k \Big] \hat{U}_\mathrm{D} + \sum_{k,d}\epsilon_{k}\hat X^\dagger_{k,d}\hat X_{k,d} \nonumber\\
&= \sum_{k,i\in \{\pm\}}\omega_{k,i}\hat{P}_{k,i}^{\dagger}\hat{P}_{k,i} + \sum_{k,d}\epsilon_{k}\hat X^\dagger_{k,d}\hat X_{k,d} 
\end{align}
where $\hat X^\dagger_{k,b} = \frac{1}{\sqrt{\mathcal{S}}}\sum_{m} \sin(k_z z_m )\hat X^\dagger_{k,m}$ are bright layer exciton operators, with the normalization constant $\mathcal{S} = \sum_m \sin^2(k_z z_m )$. Here, $\hat X^\dagger_{k,b}$  creates an exciton delocalized over all layers with an in-plane wavevector $k$. On the other hand, $\hat X^\dagger_{k,d} = \sum_{m} s_{m,d}\hat X^\dagger_{k,m}$ \added{is a dark layers exciton operator}, with $\sum_m s^*_{m,d}s_{m,d'} = \delta_{d,d'}$ \added{($\delta_{d,d'}$ is the Kronecker delta)} and $\frac{1}{\sqrt{\mathcal{S}}}\sum_m s_{m,d} \sin(k_z z_m ) = 0$. \added{Mathematically, in the absence of phonons,} these dark layers' exciton operators do not couple to cavity radiation modes, \added{and} form dark exciton bands, which are illustrated in Fig.~\ref{Fig1_schemband}d. In the last line of Eq. \ref{BFT}, we have introduced the upper and lower polariton operators, 
\begin{align}\label{P-operators}
\hat{P}_{k,+}^{\dagger} &=  \sin\theta_k \hat a^\dagger_{k} + \cos\theta_k  \hat X^\dagger_{k,b}\\
\hat{P}_{k,-}^{\dagger} &=  \cos\theta_k  \hat a^\dagger_{k}- \sin\theta_k  \hat X^\dagger_{k,b}
\end{align}
where $\theta_k = \frac{1}{2}\tan^{-1}[2\sqrt{\mathcal{S}}\Omega_k/(\omega_k - \epsilon_k)]$ is a light-matter mixing angle~\cite{Kowalewski2016JCP, MandalCR2023}. This diagonal form enables us to propagate the exciton polariton Hamiltonian in an efficient manner. Finally, expectation value of an operator $\hat{A}$ is computed as $\langle \hat{A} \rangle \approx \big\langle \langle {\Psi} (t)|\hat{A}| {\Psi} (t)\rangle \big\rangle_{\mathrm{MFE}}$ where $\langle ... \rangle_{\mathrm{MFE}}$
indicates averaging over trajectories generated from the initial realizations of nuclear coordinates $\{{q}_{n,m}(0), {p}_{n,m}(0)\}$. Details of our propagation scheme are provided in the SI.

{\bf Results and Discussion.} Here {we present results for exciton-polariton transport} where the exciton-phonon part of the system is modeled based on a multilayered perovskite material~\cite{Svenja2020FrenkelHolsteinperovskites}, which has on-site energy $\epsilon_0 = 3.2$ eV, a phonon frequency of $\omega = 1440$ cm$^{-1}$ and a hopping parameter of $\tau = 400$ cm$^{-1}$ with an interlayer spacing of $4$ nm and a lattice spacing of $a = 1.2$ nm. However, the conclusions of this work are generic and apply to a wide variety of organic and inorganic exciton-polaritons. Further details of the parameters used in this work are provided in the SI. 

Fig.~\ref{fig2_vgs}a-b present the angle-resolved polariton spectra computed using our quantum dynamical approach (see details in the SI) for a single-layer material coupled to cavity radiation modes. The theoretical polariton dispersion curves presented in Fig.~\ref{Fig1_schemband}c, which are obtained by diagonalizing $\hat{H}_\mathrm{EP}$ (see Eq.~\ref{BFT}), feature smooth polariton curves, as expected. In contrast, phonon interaction with the exciton-polaritons introduces vibronic structure to the polariton dispersion, as can be seen in  Fig.~\ref{fig2_vgs}a. Increase in the phonon coupling $\gamma = 2\gamma_0$ (where $\gamma_0 = 1.1\sqrt{\omega^3}$), in Fig.~\ref{fig2_vgs}b makes the vibronic structure more prominent. The existence of these vibronic bands and their relation to phonon coupling $\gamma$ can be, very crudely, explained due to the Franck–Condon overlaps between displaced harmonic oscillator states of the phonons. In our recent work~\cite{blackham2025}, we have presented a microscopic theory, where these successive vibronic bands describe excitations to a phonon field, which provides a quantitatively accurate description.   

Importantly, the vibronic structure in polariton dispersion is directly reflected in the group velocities obtained via direct quantum dynamics simulations.   In our quantum dynamics simulations, we model a resonant excitation in the lower polariton at the energy $E_0$ by selecting an exciton-polariton wavefunction $|\Psi (0)\rangle = \sum_{k \in \mathcal{E}_\mathrm{exc}} \tilde{c}_k \hat{P}_{k,-}^{\dagger} |\bar{0}\rangle$, which is energetically localized within an energy window $\mathcal{E}_\mathrm{exc}$ (set to be $50$ meV around $E_0$). We choose the coefficients $\tilde{c}_k$ via a Monte Carlo approach (see SI) to ensure that $|\Psi (0)\rangle$ is also spatially localized at the center in the plane of the materials (in the $\vec{e}_x$ direction). 

Fig.~\ref{fig2_vgs}c-d presents the group velocities extracted from the time-dependent excitonic density $\langle \Psi (t)| \hat{X}_{n,m}^{\dagger}\hat{X}_{n,m}|\Psi (t)\rangle$ at short times ($t < 0.3$ ps) with different numbers of layers of the material coupled to the cavity radiation. In Fig.~\ref{fig2_vgs}c we keep each layer's coupling to the radiation $\Omega_k$ a constant. As a result, the overall Rabi-splitting $\sqrt{\mathcal{S}}\Omega_k$ roughly scales with $\approx \sqrt{M}$ where $M$ is the number of layers. The group velocities obtained numerically (solid lines with filled circles)  show an oscillatory pattern due to the presence of the vibronic structure in the exciton-polariton dispersion. An increase in the number of layers, from $M = 1$ (yellow) to $M = 5$ (light red), and $M = 10$ (dark red), leads to an overall redshift due to the increase in the Rabi-splitting. Notably, the computed group velocities are much smaller (renormalized) than the theoretical group velocity $\frac{d}{dk} \omega_{-,k}$ (i.e. slope of the dispersion) in bare exciton-polariton. This renormalization effect has been observed in multiple recent experiments~\cite{XuNC2023, BalasubrahmaniyamNM2023,PandyaAS2022}. To clearly analyze the role of multiple layers in the exciton-polariton dynamics, in Fig.~\ref{fig2_vgs}d and the rest of the results presented in this work we keep $\sqrt{\mathcal{S}}\Omega_k$ a constant.

Fig.~\ref{fig2_vgs}d presents the exciton-polariton group velocities in a multilayered material (with $M = 25$) compared to a single-layered material at various phonon couplings. Fig.~\ref{fig2_vgs}d reveals that for the same effective Rabi splitting, $\sqrt{\mathcal{S}}\Omega_k$, the exciton-polariton group velocity in a multilayered material could be substantially higher than the single-layered material. This is further illustrated in Fig.~\ref{fig2_vgs}e which presents the enhancement of group velocity as a function of the number of layers placed inside an optical cavity. We find that the exciton-polariton density in multilayered material can propagate with velocities upwards of $\approx$25\%  higher than in a single-layer material. 
 
This is surprising because the number of phonon modes interacting with the exciton-polariton subsystem is significantly higher in a multilayered material compared to a single-layer material. However, despite the increase in the number of phonon modes, which one expects to provide additional sources of dissipation and fluctuation, the exciton-polariton propagates faster with a longer decoherence time.  We find that this effect originates from a synchronization of phonon fluctuations in the collective bright layer, leading to the suppression of phonon-induced decoherence and enhanced group velocity. 
 
{\bf Phonon fluctuation synchronization effect.} To provide a microscopic understanding of \added{phonon fluctuation synchronization effect},  we analyze the term $\hat{H}_\mathrm{ex-phn}$ in the light-matter Hamiltonian. We introduced the bright exciton operators in real space as  $\hat X^\dagger_{n,b} = \frac{1}{\sqrt{\mathcal{S}}}\sum_{m} \sin(k_z z_m )\hat X^\dagger_{n,m} = \frac{1}{\sqrt{N}}\sum_n \hat X^\dagger_{k,b} e^{-ik x_n}$. Since only the bright exciton operators $\{\hat X^\dagger_{n,b}\} \equiv \{\hat X^\dagger_{k,b}\}$ directly contribute (see Eq.~\ref{BFT}) in the dynamics of the exciton-polaritons, we focus on the  exciton-phonon coupling term in the subspace defined by the projection operator $\mathcal{P} = \sum_{n} \hat X^\dagger_{n,b} \hat X_{n,b} =  \sum_{n} |n,b\rangle \langle n,b|$, which is written as
\begin{align}\label{exc-phonon2}
\mathcal{P} \left( \hat H_\mathrm{ex-phn} \right)\mathcal{P}  &= \gamma \sum_{n}   \hat X^\dagger_{n,b}\hat X_{n,b} \left( \frac{1}{{\mathcal{S}}}\sum_m {q}_{n,m} \sin^2(k_z z_m)\right) \nonumber \\
&=  \gamma \sum_{n}   \hat X^\dagger_{n,b}\hat X_{n,b} \tilde{q}_{n}, 
\end{align} 
where $\tilde{q}_{n} = \frac{1}{{\mathcal{S}}}\sum_m {q}_{n,m} \sin^2(k_z z_m)$. \added{Note that we do not make such assumption when performing any numerical calculations.} Therefore, exciton-polariton dynamics effectively interact with the collective phonon fluctuations $\tilde{q}_{n}$ that are delocalized over all the layers. Importantly, the $\tilde{q}_{n}$ distribution is much narrower than the distribution of ${q}_{n,m}$, effectively exhibiting a phonon fluctuation synchronization in multilayered materials. This phonon fluctuation synchronization \added{effect} can also be viewed as an effective reduction in temperature (shrinking the thermal distribution of the phonons), leading to enhanced coherent transport of exciton-polaritons (see the SI for more details). 

Fig.\ref{fig2_vgs}f presents the distribution of the excitonic on-site energy \added{in single layer materials} compared to the multilayered materials. Due to the \added{phonon} fluctuation synchronization \added{effect} discussed above, the fluctuations in the effective excitonic on-site energy $\tilde {\epsilon}_n = {\epsilon}_0 + \gamma \tilde{q}_{n}$ in the collective bright layer {of multilayer materials} is much less compared to the on-site energies ${\epsilon}_n = {\epsilon}_0 + \gamma {q}_{n}$  in the single-layer scenario. As a result, the exciton-polariton propagation is enhanced as \added{the exciton} hopping between neighboring sites is easier, which is schematically illustrated in Fig.\ref{fig2_vgs}g-h. 

To test our phonon fluctuation synchronization hypothesis, in Fig.~\ref{fig2_vgs}i, we performed \added{mixed-quantum-classical} simulations in multilayered materials by sampling the phonons in a constrained fashion, i.e. ${q}_{n,m}(t = 0) = {q}_{n,1}(t = 0)$, such that the initial position of a phonon at a site $n$ is the same regardless of the layer \added{index}. As a result, $\tilde{q}_{n} = {q}_{n,1}$, which means that the fluctuations experienced by the collective bright layer in a multilayered material are the same as in the single-layer material.  The group velocities presented in Fig.~\ref{fig2_vgs}i show that the dynamics of the multilayered material under constrained sampling are identical to the single-layer material, corroborating our phonon fluctuation synchronization hypothesis. 

Fig.~\ref{Fig3_spread} presents time-dependent exciton-polariton density (exciton population at each site $n$) to illustrate how the phonon fluctuation synchronization in multilayered materials can enable coherent transport. Fig.~\ref{Fig3_spread}a presents the time-dependent exciton\added{-polariton} population in the absence of phonon ($\gamma = 0$) \added{for both single-layer and multilayer materials}. As expected, the exciton-polariton population, initially prepared at the center, propagates ballistically. In the presence of strong phonon interaction ($\gamma \neq 0$) the exciton-polariton propagation in a single-layer material, presented in Fig.~\ref{Fig3_spread}b, appears incoherent, exhibiting diffusive transport. In contrast,
the exciton-polariton transport in multilayered material at the same exciton-phonon coupling $\gamma$, presented in Fig.~\ref{Fig3_spread}c, shows ballistic coherent propagation of the exciton-polariton wavefront. This remarkable effect can be clearly seen in  Fig.~\ref{Fig3_spread}d, which presents the exciton-polariton density at $t = 0.75$ ps. We see that the wavefront in a multilayered material propagates coherently, which is missing in the single-layer case. 

Fig.~\ref{Fig3_spread}g-h further illustrates that the exciton-polariton wavefront is delocalized over all the layers, with Fig.~\ref{Fig3_spread}e-f presenting the corresponding results in the single-layer scenario. This shows how delocalization of exciton-polariton density in the transverse direction exhibits enhanced transport in the longitudinal direction (or in the plane of the material). We find that this phonon fluctuation synchronization effect is quite ubiquitous, showing up in a wide range of parameter regimes. We present more exciton-polariton density propagation data for different values of $E_0$, $\Omega_0$, and $\gamma$ in the SI.

\begin{figure}[!h]
    \centering
    \includegraphics[width=1\linewidth]{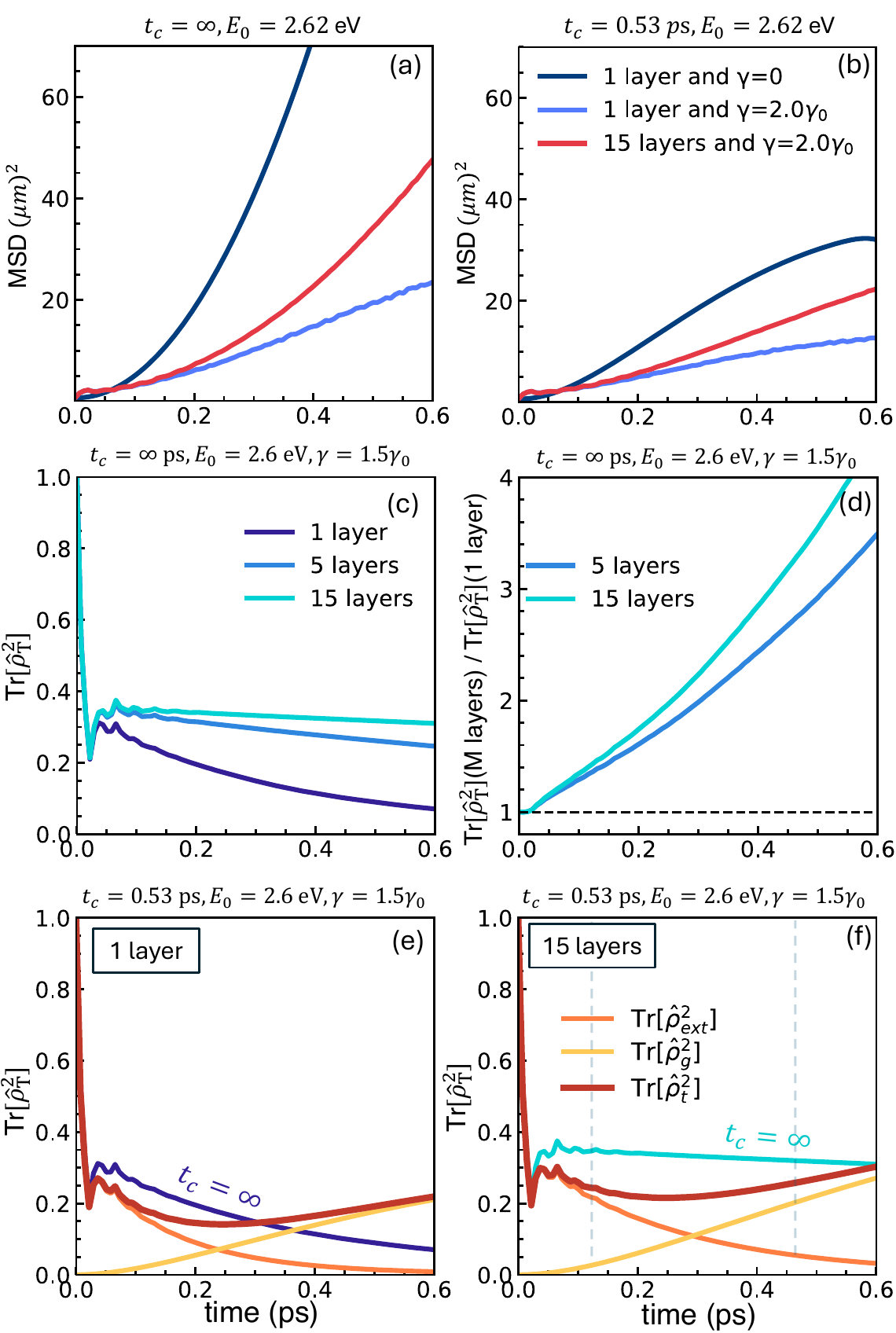}
    \caption{\footnotesize \textbf{Effect of cavity photon loss on mean-square displacement (MSD) and purity in multilayered materials \protect \added{in a two-dimensional cavity}.} MSD in single-layer material vs. multilayered material \protect \added{($M=15$) with initial energy of $E_0=2.62$ eV} (a) in the absence ($t_c=\infty$) and (b) presence \protect \added{($t_c=0.53$ ps)} of cavity photon loss. (c) \protect \added{Purity for different numbers of layers (M) and the initial excitation at \protect \added{$E_0=2.6$ eV} in the absence of photon loss, (d) the ratio of the purity of multilayered materials ($M>1$) relative to the purity of a single-layer material ($M=1$) with the initial excitation at $E_0=2.6$ eV in the absence of photon loss}.
    Total purity (Tr$[\hat \rho_\text{T}^2(t)]$) along with exciton-polariton (Tr$[\hat \rho_\text{EP}^2(t)]$) and ground state (Tr$[\hat \rho_\text{g}^2(t)]$) contributions to total purity in the presence of cavity photon loss (\protect \added{$t_c=0.53$ ps}) and initial excitation at \protect \added{$E_0=2.6$ eV} for (e) single layer and (f) 15 layers.  ($\gamma_0\approx 5.8\times 10^{-4}$ a.u., \protect \added{$\Omega_0 = 330$ meV, and 40001 sites.}) } 
    \label{Fig4_loss}
\end{figure}

\begin{figure*}[t]
    \centering    \includegraphics[width=0.75\linewidth]{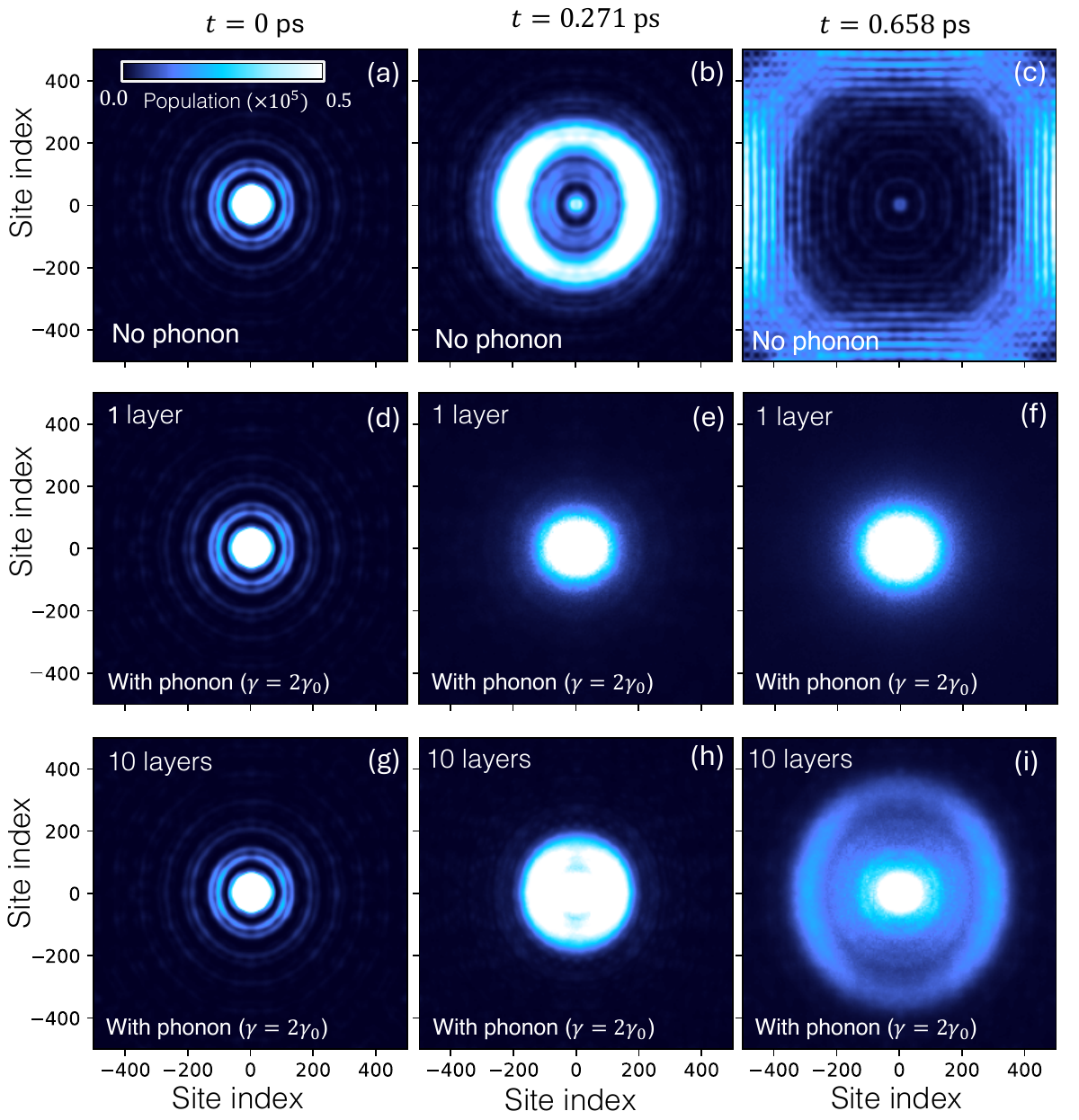}
    \caption{\footnotesize \protect\added{\textbf{Comparing exciton-polariton dynamics in single-layer vs. multilayered materials ($M=10$)  with the initial excitation at $E_0 = 2.43$ eV in a three-dimensional cavity.} (a)-(c) Time-dependent exciton-polariton population (density) in the absence of exciton-phonon coupling for both single-layer and multilayer materials at three different times, (d)-(f) same as (a)-(c) but for the single-layer material and in the presence of exciton-phonon coupling ($\gamma=2\gamma_0$), (g)-(i) same as (a)-(c) but for the multilayered material ($M=10$) and in the presence of exciton-phonon coupling ($\gamma=2\gamma_0$). To determine the multilayer exciton-polariton populations, we summed the density contributions from all layers at each site. Population for longer times for all cases can be found in SI. ($\gamma_0\approx 5.8\times 10^{-4}$ a.u., $\Omega_0 = 330$ meV, and $N_x=N_y=1000$.)} } 
    \label{Fig5_3Dcavity}
\end{figure*}

Fig.~\ref{Fig4_loss}a-b presents the excitonic mean square displacement (MSD) of the exciton-polariton \added{initially prepared} at an energy window of \added{$2.62\pm0.025$ eV}. The excitonic MSD is computed as~\cite{XuNC2023, TichauerAS2023, ChngNL2025} 

\begin{align}\label{msd}
\mathrm{MSD} = a^2 \left(\sum_n n^2  P_n(t) - \left(\sum_n n P_n(t)\right)^2
\right),
\end{align}
where $P_n(t) = \left\langle\frac{1}{{\mathcal{N}}}\langle\Psi (t)| \hat X^{\dagger}_n \hat X_n|\Psi (t) \rangle \right\rangle_\mathrm{MFE}$ is excitonic density at the site $n$ with $\mathcal{N} = \sum_n \langle\Psi (t)| \hat X^{\dagger}_n \hat X_n|\Psi (t) \rangle $ a normalization constant. In the absence of phonons ($\gamma = 0$), MSD is quadratic in time (dark blue line in Fig.~\ref{Fig4_loss}a), i.e. ballistic, as expected. In the presence of exciton-phonon coupling ($\gamma \neq 0$), MSD is quadratic \added{in time} till $\approx 0.3$ ps and becomes almost \added{a linear function of time} beyond this time, thereby exhibiting diffusive transport as phonon-induced decoherence sets in. For the same exciton-phonon coupling, the MSD appears to be ballistic for a longer time in the case of multilayered material, suggestive of extended coherence lifetime due to the phonon fluctuation synchronization \added{effect} discussed above. Thereby, the MSD reaches a higher value in the case of a multilayered material compared to a single-layer material.

Fig.~\ref{Fig4_loss}b presents MSD in the presence of cavity loss that accounts for the finite lifetime of a photon trapped inside an imperfect optical cavity.  Overall, the trends are the same as in Fig.~\ref{Fig4_loss}a but with an overall lower MSD. These results agree with recent theoretical and experimental works, which demonstrate that MSD is suppressed in the presence of cavity loss~\cite{ChngNL2025, TichauerAS2023, PandyaAS2022}.

While ballistic transport is suggestive of coherent dynamics, it only provides indirect evidence of decoherence suppression. To provide a quantitative measure and a direct confirmation of decoherence suppression, we compute the purity~\cite{Wenxiang2022JPCL, Arnab2017JCP, Franco2013JCP}, which is defined as $\mathrm{Tr}[\hat{\rho}_\mathrm{EP}^2(t)]$ where 
$\hat{\rho}_\mathrm{EP}(t)$ is the reduced density matrix describing the exciton-polariton subsystem. Within our mixed-quantum classical simulations, the exciton-polariton reduced density matrix is computed as $\big\langle |\Psi (t) \rangle\langle \Psi (t) |\big\rangle_\mathrm{MFE}$. In the SI, we provide an efficient approach to computing the purity in our mixed quantum-classical simulations. 

\added{Fig.~\ref{Fig4_loss}c} presents the time-dependent purity $\mathrm{Tr}[\hat{\rho}_\mathrm{EP}^2(t)]$ for initial excitation of \added{$E_0 = 2.6$ eV}. In the single-layer scenario (violet solid line), at $t=0$, the purity is one, indicating a pure (coherent) state. At longer times, $ t \approx 0.6$ ps, the purity decays toward zero, indicating phonon-induced decoherence of the exciton-polariton subsystem. \added{Fig.~\ref{Fig4_loss}c}, shows that an increase in the number of layers significantly suppresses the decay of the purity, directly depicting an extended coherence lifetime in multilayered materials. Overall, we find that multilayered materials can enhance purity by up to a factor of \added{$\approx 4$} as shown in Fig.~\ref{Fig4_loss}d, which presents the ratio of purity in a multilayered material \added{relative} to a single-layer material. 

Fig.~\ref{Fig4_loss}e-f presents the time-dependent purity in the presence of cavity loss. Note that in the presence of cavity loss, the population in the exciton-photon subspace $\{|1_k\rangle, |n,m\rangle  \}$  leaks out to the ground state $|\bar{0}\rangle$. As a result, the total reduced density matrix (when tracing over the phonon degree of freedom) can be defined as $\hat{\rho}_\mathrm{T}(t) = \hat{\rho}_\mathrm{g}(t) + \hat{\rho}_\mathrm{EP}(t)=  \rho_\mathrm{gg}(t) |\bar{0}\rangle\langle \bar{0}| + \sum_{\alpha,\beta}\rho_{\alpha\beta}(t) |\alpha\rangle\langle \beta|$ where $\{|\alpha\rangle, |\beta\rangle\} \in \{|1_k\rangle, |n,m\rangle  \}$. As a result, the total purity can be written down as
\begin{align}
  \mathrm{Tr}[\hat{\rho}_\mathrm{T}^2(t)] = \mathrm{Tr}[\hat{\rho}_\mathrm{g}^2(t)] +  \mathrm{Tr}[\hat{\rho}_\mathrm{EP}^2(t)] . 
\end{align}

These three quantities, $\mathrm{Tr}[\hat{\rho}_\mathrm{T}^2(t)]$, $\mathrm{Tr}[\hat{\rho}_\mathrm{g}^2(t)] = \rho_\mathrm{gg}^2(t)$, and  $\mathrm{Tr}[\hat{\rho}_\mathrm{EP}^2(t)]$, are presented in Fig.~\ref{Fig4_loss}e-f at a cavity photon lifetime $t_c  = 0.53$ ps. Note that at $t \gg t_c $ the total purity reaches 1 because the total density matrix $\hat{\rho}_\mathrm{T}(t\gg t_c) \rightarrow |\bar{0}\rangle \langle \bar{0}|$ which is a pure state. At the same time, the coherence in the exciton-polariton subspace becomes zero with $\hat{\rho}_\mathrm{EP} \rightarrow 0$. Overall, the total purity (red solid line) decays till $\approx 0.2$ ps primarily due to the decay of exciton polariton purity $\mathrm{Tr}[\hat{\rho}_\mathrm{EP}^2(t)]$. The rise in the total purity after $0.3$ ps is due to the increase in $\rho_\mathrm{gg}^2(t)$ as the population leaks to the ground state. Compared to the no cavity loss scenario, represented by the violet and cyan solid line in Fig.~\ref{Fig4_loss}e-f respectively, the $\mathrm{Tr}[\hat{\rho}_\mathrm{EP}^2(t)]$ decays faster, which is due to the loss in the total population in the exciton-polariton subspace. Importantly, regardless of the presence or absence of cavity loss, the multilayered material suppresses the decay in exciton-polariton purity. This is expected as the phonon fluctuation synchronization effect in the multilayered materials only protects against phonon-induced decoherence and does not directly combat cavity loss.

{{\bf Three-dimensional simulations.} 
\added{Here we demonstrate that the phonon fluctuation synchronization effect on the exciton-polariton dynamics is not limited to a simplified two dimensional light-matter system, and the same effect exists when considering the full three-dimensional light-matter Hamiltonian. For this, we consider two mirrors (DBR mirrors) placed in the $xy$-plane which are located at $z=0$ and $z=L$. For this cavity setup, we still consider the first cavity quantization mode along the $z$ direction such that $k_z = \pi/L$, therefore $\boldsymbol{k} = \boldsymbol{k}_\parallel + k_z\vec{e}_z = k_x\vec{e}_x + k_y\vec{e}_y + k_z\vec{e}_z$ for  $\boldsymbol{k}_\parallel$ being the projection of the wavevector onto $xy$-plane. According to the solutions of Maxwell equations in three-dimensions for a Fabry-Pérot cavity, the cavity photons at each wavevector $\boldsymbol{k}$ consist of two cavity polarizations: the transverse electric field (TE or $s$) mode, and transverse magnetic field (TM or $p$) mode, which are degenerate for a bare cavity. For the bare exciton part of the light-matter Hamiltonian, we consider two-dimensional stacked multilayered material placed in parallel to the cavity mirrors. When considering a single-layer material, we place them at the middle of the optical cavity. For  multilayered materials, additional layers are stacked above and below a layer placed in the middle of the optical cavity with an interlayer spacing of $a_z = 4~\text{nm}$. The excitonic part of the system is described by a two-dimensional tight-binding model on a rectangular lattice with nearest-neighbor hopping. The lattice consists of $N_x$ sites along the $x$ direction with spacing $a_x$, and $N_y$ sites along the $y$ direction with spacing $a_y$. }

\added{Considering one exciton per unit cell, with a transition dipole aligned to the $y$ direction, the form of the light-matter interaction beyond the long-wavelength approximation is given by
\begin{equation}
\begin{split}
&\hat{H}_\text{ex-cav} =\\
&\frac{\Omega_0}{\sqrt{N}}\sum_{\bar{n},\boldsymbol{k}}\hat{X}^\dagger_{\bar{n}}\left(   \sqrt{\frac{\omega_{\boldsymbol{k}}}{\omega_0}} {k_y} 
\hat{a}_{\boldsymbol{k}} - i \sqrt{\frac{\omega_0}{\omega_{\boldsymbol{k}}}} k_x \hat b_{\boldsymbol{k}}\right)\frac{\sin(k_z z_m)}{|\boldsymbol{k}_\parallel|}e^{i\boldsymbol{k}_\parallel \cdot \boldsymbol{R}_{\bar{n}} }+h.c.
\end{split}
\end{equation}
where $N=N_xN_y$ and $\hat{X}^\dagger_{\bar{n}}$ creates exciton at site $\boldsymbol{R}_{\bar{n}}=\boldsymbol{R}_{n_x,n_y, m}=n_xa_x\vec{e}_x+n_ya_y\vec{e}_y+ma_z\vec{e}_z$ for  $n_x\in [0, N_x-1]$, $n_y\in [0, N_y-1]$, and $m$ is the layer index. Photon frequencies in this three-dimensional setup is given by $\omega_{\boldsymbol{k}}=\frac{c}{\eta}\sqrt{|\boldsymbol{k}_\parallel|^2 + k_z^2}$ for $\eta=2.4$. Details of the light-matter Hamiltonian in a three-dimensional light-matter Hamiltonian are provided in the SI.}

\added{Fig. \ref{Fig5_3Dcavity} presents the temporal exciton-polariton real-space population (density) in a three-dimensional optical cavity. For this figure, we use the exciton lattice constants of $a_x = a_y=20~\text{nm}$, $N_x=N_y=1000$, and all other parameters are the same as in Fig.~\ref{fig2_vgs}-\ref{Fig4_loss}.
For the exciton-photon initial state, we model a resonant excitation in the lowest polariton band (defined by $ \hat{P}_{k,0}^{\dagger}$ operator) at the energy $E_0$ by selecting an exciton-polariton wavefunction $|\Psi (0)\rangle = \sum_{k \in \mathcal{E}_\mathrm{exc}} \tilde{c}_k \hat{P}_{k,0}^{\dagger} |\bar{0}\rangle$, which is energetically localized within an energy window $\mathcal{E}_\mathrm{exc}$ (set to be $40$ meV around $E_0$). Details of the initial state and the split-operator approach for the full three-dimensional light-matter system is provided in the SI.}

\added{Fig.\ref{Fig5_3Dcavity}a-c shows ballistic coherent propagation of exciton-polariton population in the absence of phonon ($\gamma=0$). Fig.\ref{Fig5_3Dcavity}d-f demonstrates that for the single-layer material in the presence of exciton-phonon coupling ($\gamma=2\gamma_0$), exciton-polariton propagation is diffusive and incoherent, and the exciton-polariton population is localized near the center of the optical cavity where our initial state was prepared. However, the exciton-polaritonic dynamics in a multilayered material with same exciton-phonon coupling ($\gamma=2\gamma_0$) is ballistic and coherent as it is shown in Fig.\ref{Fig5_3Dcavity}g-i. For a multilayer material, the propagation of the exciton-polariton population wavefront remains coherent beyond $t=0.658~\text{ps}$, which is presented in the SI. These results mirror the simpler two-dimensional scenario presented in Fig.~\ref{Fig3_spread}, validating  our understanding of phonon fluctuation synchronization effect in three-dimensions.}

{\bf Conclusion.} In this work, we have studied the exciton-polariton dynamics in multilayer materials and compared their transport and coherence to single-layer materials. To achieve this, we introduced a new, fast, and highly efficient mixed quantum-classical approach that utilizes a {\it bright layer} description of the light-matter interactions, allowing us to simulate a large number of quantum states \added{($\approx 10^{10}$)}. 

Using our quantum dynamics approach, we found a phonon fluctuation synchronization effect where the delocalized light-matter interactions in a multilayered material shield exciton-polaritons from phonon-induced decoherence. As a result, exciton-polaritons propagate faster with an increased group velocity and, {\it more importantly}, remain more coherent for a longer time in multilayered materials despite interacting with a significantly larger number of phonon modes. We show that this phonon fluctuation synchronization effect originates from a fluctuation averaging in the transverse direction (perpendicular to the mirror plane)  over multiple layers that suppresses dynamical disorder in the longitudinal direction (in the plane of the materials/mirrors), enhancing transport in the material plane. \added{Mathematically we show} that \added{under an appropriate unitary transformation,} only one effective layer (the bright layer) couples with the cavity modes, while the rest remain uncoupled (dark layers). \added{Although, the phonons can transfer population between these dark layers and the bright layer or the cavity modes}. It is this bright layer, delocalized over all layers, that experiences fewer phonon fluctuations than in a single-layer material. We also show that even in the presence of cavity loss, multilayered material enhances transport and coherence, although to a lesser extent. 

Our work provides microscopic insights into exciton-polariton dynamics in multilayered materials and illustrates that quantum coherence lifetime can be extended by using a multilayered cavity architecture. This work reveals that the pathway to polariton quantum devices is likely via stacked multilayered materials.

\section { Data Availability}
    The data that support the plots within this paper and other findings of this study are available from the corresponding authors upon a reasonable request.
\section { Code Availability}

The source code that supports the findings of this study is available from the corresponding author upon reasonable request.

\section { Acknowledgments}
This work was supported by the Texas A\&M startup funds. This work used TAMU FASTER at the Texas A\&M University through allocations  PHY230021 and PHY240260 from the Advanced Cyberinfrastructure Coordination Ecosystem: Services \& Support (ACCESS) program, which is supported by National Science Foundation grants \#2138259, \#2138286, \#2138307, \#2137603, and \#2138296. A.M. appreciates inspiring discussions with Yi Rao, Milan Delor, Daniel Tabor, Dong Hee Son, Pengfei Huo, David R. Reichman, Pritha Ghosh, and Sachith Wickramasinghe.  
 
\section{ Author contributions}
S. R. K. and A. M. designed the research. S. R. K., L. B. and A. M. (Arshath Manjalingal) developed the quantum dynamical approach.  and analyzed data. S. R. K. performed the quantum dynamical simulations.  S. R. K., L. B. and A. M. (Arshath Manjalingal) and A. M. analyzed data and wrote the manuscript.

\section{ Competing Interests}
The authors declare no competing interests.

}
\bibliography{bib.bib}
\bibliographystyle{naturemag}

\end{document}